# On Enforcing Satisfiable, Coherent, and Minimal Sets of Dyadic Relation Constraints in *MatBase*

**Christian Mancas***

*Mathematics & Computer Science Department, Ovidius University at Constanta, Romania*

***Corresponding Author:** Christian Mancas, Mathematics & Computer Science Department, Ovidius University, Bd. Pipera 1/U, Voluntari, IF, Romania.

### Abstract

This paper rigorously and concisely defines, in the context of our (Elementary) Mathematical Data Model ((E)MDM), the mathematical concepts of dyadic relation, reflexivity, irreflexivity, symmetry, asymmetry, transitivity, intransitivity, Euclideanity, inEuclideanity, equivalence, acyclicity, connectivity, the properties that relate them, and the corresponding corollaries on the coherence and minimality of sets made of such dyadic relation properties viewed as database constraints. Its main contribution is the pseudocode algorithm used by *MatBase*, our intelligent database management system prototype based on both (E)MDM, the relational, and the entity-relationship data models, for enforcing dyadic relation constraint sets. We proved that this algorithm guarantees the satisfiability, coherence, and minimality of such sets, while being very fast, solid, complete, and minimal. In the sequel, we also presented the relevant *MatBase* user interface as well as the tables of its metacatalog used by this algorithm.

### Abbreviations

(E)MDM = (Elementary) Mathematical Data Model.
DBMS = Database Management System.
db(s) = database(s).
iff = if and only if.

### Introduction

We presented in [1] the current version of our (Elementary) Mathematical Data Model ((E)MDM). Out of its 76 constraint types, there are 11 pertaining to dyadic relations: reflexivity, irreflexivity, symmetry, asymmetry, transitivity, intransitivity, Euclideanity, inEuclideanity, equivalence, connectivity, and acyclicity. As usual in mathematics, some of them or some combinations of them imply others, while some of them are mutually exclusive. This is why any intelligent Database Management System (DBMS) must accept only satisfiable, coherent, and, for optimality concerns, also minimal sets of constraints.

*MatBase* [2] is our intelligent DBMS prototype, based on both (E)MDM, the Entity-Relationship Data Model [3-5], the Relational Data Model [5-7], and Datalog¬ [7, 8], currently implemented in two MS platforms: Access (for small dbs and undergraduate students) and .NET C# and SQL Server (for large dbs and MSc. students). Its (E)MDM interface provides users with a form (see, e.g., Figure 1) in which all metadata [9] for any set of a database (db) it manages may be inspected and updated.

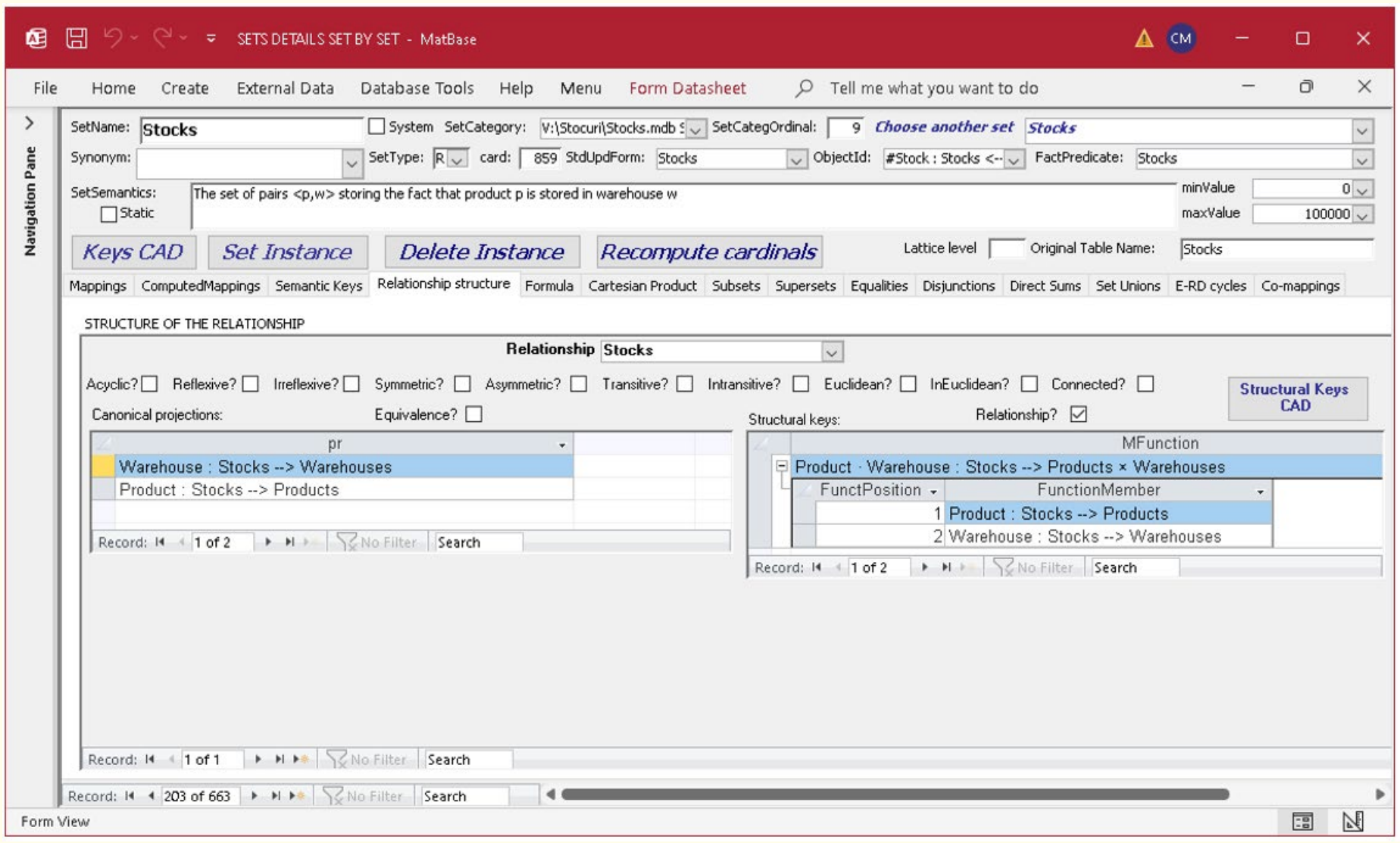

***Figure 1:*** MS Access *MatBase* form for managing db sets' schema.

In particular, for dyadic relationships users may assert or delete their properties by simply clicking on the corresponding checkboxes from the *Relationship structure* tab. Immediately after each such click, *MatBase* analyzes the new desired such constraint set and undoes the update if it is invalid (e.g., the current relation is not a dyadic one, or the corresponding constraint set would be incoherent, or the user tried to delete a redundant constraint, or the current db instance does not satisfy the newly desired constraint set, etc.). If the update is valid, then *MatBase* not only accepts it, but also automatically updates the subset of corresponding redundant constraints and generates or deletes the code needed to enforce the newly desired dyadic relation type constraint set.

This paper describes the math behind this process, as well as the metadata and algorithm that *MatBase* uses to perform these tasks. Of course, that these 11 constraint types are non-relational, i.e., they may not be enforced by any relational DBMS (e.g., MS SQL Server, Oracle Database, IBM DB2, etc.). Consequently, they should be enforced by db software applications managing the corresponding relational dbs. *MatBase* automatically generates such software applications for every db it manages.

***Related work***

*MatBase's* constraint sets coherence and minimality enforcement algorithms were generally presented at a higher conceptual level in [10]. First, [10] deals with all (E)MDM constraint types (which were only 61 at that time); then, it does not address the particularities of dyadic relations, which are cases of homogeneous binary function products (i.e., of type $f \bullet g : D \rightarrow (C \cup \text{NULLS})^2$), for which both functions are totally defined (i.e., they may not take null values) and their product is minimally one-to-one (as dyadic relations are sets, so they do not allow for duplicates). Moreover, [10] does not deal with the constraint sets which imply the universality of the corresponding dyadic relations.

Deeper details on dyadic relationship enforcement in *MatBase* were presented in [11, 12].

Proofs of the mathematical results presented in the next section may be found, e.g., in [8, 13].

(E)MDM is also a 5th generation programming language [14, 15] and *MatBase* is also a tool for transparent programming while modeling data at conceptual levels [2].

To our knowledge, the other most closely related approaches to non-relational constraint enforcement are based on business rules management (BRM) [16, 17] and their corresponding implemented systems (BRMS) and process managers (BPM), like the IBM Operational Decision Manager [18], IBM Business Process Manager [19], Red Hat Decision Manager [20], Agiloft Custom Workflow/BPM [21], etc. They are generally based on XML (but also on the Z notation, Business Process Execution Language, Business Process Modeling Notation, Decision Model and Notation, or the Semantics of Business Vocabulary and Business Rules), which is the only other field of endeavor trying to systematically deal with business rules, even if informally, not at the db design level but at the software application one, and without providing automatic code generation.

From this perspective, (E)MDM is also a BRM but a formal one, and *MatBase* is also a BRMS but an automatically code generating one.

The satisfiability, coherence, and minimality of first order predicate formulae sets has been extensively studied mathematically (e.g., [22]) but not in the db contexts, as there are only six relational constraint types (for which any combination is coherent), out of which NoSQL DBMSes only use 2 or 3.

## Materials and Methods

The following definitions, propositions, and corollaries are from Appendix A ("The Math Behind (E)MDM") of [8]. The propositions are from its subsection A.3.1.3 ("Relations"), while the corollaries from its subsection A.5.2 ("Coherence and Minimality of Dyadic Relationship Constraint Sets"), where they are numbered from A.5.2.1 to A.5.2.19.

### *Definitions*

1. A *dyadic relation* $R$ is a subset of a Cartesian product of a set $S$ with itself: $R \subseteq S \times S$; iff $R = S \times S = S^2$ then it is said to be *universal*.
2. A dyadic relation $R$ over a set $S$ (having any distinct elements $x, y, z$) is:
   a. *reflexive* iff $xRx$.
   b. *irreflexive* iff $\neg(xRx)$.
   c. *symmetric* iff $xRy$ then $yRx$.
   d. *asymmetric* iff $xRy$ then $\neg(yRx)$.
   e. *transitive* iff $xRy$ and $yRz$ then $xRz$.
   f. *intransitive* iff $xRy$ and $yRz$ then $\neg(xRz)$.
   g. *Euclidean* iff $xRy$ and $xRz$ (right-Euclideanity) or $yRx$ and $zRx$ (left-Euclideanity) then $yRz$ and $zRy$ (i.e., both left- and right-Euclidean).
   h. *inEuclidean* iff $xRy$ and $xRz$ or $yRx$ and $zRx$ then $\neg(yRz)$ and $\neg(zRy)$ (i.e., neither left-, nor right-Euclidean).
   i. *equivalence* iff it is both reflexive, symmetric, and transitive.
   j. *connected* iff $xRy$ or/and $yRx$.
   k. *acyclic* iff $x_1Rx_2, x_2Rx_3, \ldots, x_{n-1}Rx_n$ implies $\neg(x_nRx_1)$, for any natural $n > 0$ and distinct $x_1, \ldots, x_n \in S$.
3. A *constraint* is a first order logic formula that has all its variable occurrences bound to a universal quantifier (i.e., ∀ -for any- and ∃ - there is). For example, all above 11 properties are constraint types of dyadic relations.
4. A constraint is *satisfied* by a set of values for its variables if it has value *true* for them; otherwise, it is *violated*. A constraint set is *satisfied* by a set of values for all its variables if all its constraints are satisfied.
5. A constraint set is *incoherent* iff it is satisfied only by the corresponding empty set. For example, according to the first order logic laws of *non-contradiction* ("nothing can be both true and false simultaneously") and *excluded middle* ("everything is either true or false, but not neither"), the sets {$R$ reflexive, $R$ irreflexive} and {$R$ symmetric, $R$ asymmetric} are incoherent, for any non-void dyadic relation *R*.

6. A constraint set *Γ* *implies* a constraint *c* iff *c* is *true* whenever all constraints of *Γ* are *true*. For example, as acyclicity implies irreflexivity for any dyadic relation *R* (as any *xRx* corresponds to a cycle of length 0), the set {*R* acyclic} implies the constraint *R* irreflexive.
7. A constraint *c* is *redundant* in a constraint set *Γ* iff {*Γ* - *c*} implies *c*. For example, in the set {*R* acyclic, *R* irreflexive}, *R* irreflexive is redundant.
8. A constraint set is *minimal* iff it does not contain any redundant constraint.

Obviously, any DBMS must accept only satisfiable and coherent set of constraints and should enforce only minimal ones. Moreover, universal relations should never be stored as such, but only as queries/views computable according to the corresponding Cartesian products.

In what follows, we consider any finite set *S* having at least 4 elements (which is a norm in dbs) and any dyadic non-void relation *R* $\subseteq S^2$ over it; the following propositions and corollaries hold:

***Propositions***

1. (i) (*R* reflexive ⇒ ¬(*R* irreflexive)) ∧ (*R* irreflexive ⇒ ¬(*R* reflexive)).
   (ii) (*R* symmetric ⇒ ¬(*R* asymmetric)) ∧ (*R* asymmetric ⇒ ¬(*R* symmetric)).
2. (i) *R* asymmetric ⇒ *R* irreflexive.
   (ii) *R* asymmetric ⇒ ¬(*R* Euclidean).
   (iii) *R* transitive ⇒ ¬(*R* inEuclidean).
   (iv) *R* intransitive ⇒ *R* irreflexive.
   (v) *R* intransitive ⇒ ¬(*R* Euclidean).
   (vi) *R* inEuclidean ⇒ *R* irreflexive.
   (vii) *R* connected ⇒ ¬(*R* intransitive).
   (viii) *R* connected ⇒ ¬(*R* inEuclidean).
3. *R* acyclic ⇒ *R* asymmetric ∧ *R* inEuclidean.
4. *R* Euclidean ⇔ *R* symmetric ∧ *R* transitive.
5. (i) *R* is both transitive and intransitive iff there are no distinct elements *x, y, z* ∈ *S* such that *xRy* ∧ *yRz.*
   (ii) *R* transitive ∧ *R* intransitive ⇒ *R* inEuclidean.
   (iii) *R* transitive ∧ *R* intransitive ⇒ ¬(*R* connected).
6. (i) *R* is both Euclidean and inEuclidean iff there are no distinct elements *x, y, z* ∈ *S* such that *xRy* ∧ *xRz* or such that *yRx* ∧ *zRx.*
   (ii) *R* Euclidean ∧ *R* inEuclidean ⇒ *R* intransitive.
7. *R* irreflexive ∧ *R* transitive ⇒ *R* asymmetric.
8. *R* symmetric ∧ *R* intransitive ⇒ *R* inEuclidean.
9. (i) *R* symmetric ∧ *R* inEuclidean ⇒ *R* intransitive.
   (ii) *R* symmetric ∧ *R* inEuclidean ⇒ ¬(*R* connected).
10. *R* asymmetric ∧ *R* transitive ⇒ *R* acyclic.
11. (i) *R* symmetric ∧ *R* connected ⇒ *R* Euclidean.
    (ii) *R* symmetric ∧ *R* connected ⇒ ¬(*R* intransitive).
12. (i) *R* Euclidean ∧ ¬(*R* inEuclidean) ⇒ ¬(*R* intransitive).
    (ii) *R* Euclidean ∧ ¬(*R* inEuclidean) ⇒ ¬(*R* acyclic).
    (iii) *R* inEuclidean ∧ ¬(*R* Euclidean) ⇒ ¬(*R* connected).
13. *R* transitive ∧ *R* inEuclidean ⇒ ¬(*R* connected).
14. *R* intransitive ∧ *R* Euclidean ⇒ *R* inEuclidean.

15. $R$ intransitive ∧ $R$ inEuclidean ⇒ ¬($R$ connected).
16. $R$ acyclic ∧ $R$ connected ⇒ $R$ transitive.
17. $R$ intransitive ∧ $R$ acyclic ⇒ ¬ ($R$ connected).
18. $R$ inEuclidean ∧ $R$ connected ⇒ $R$ acyclic.
19. $R$ acyclic ∧ $R$ connected ⇒ $R$ inEuclidean.
20. $R$ connected ∧ $R$ reflexive ∧ ($R$ symmetric ∨ $R$ Euclidean) ∨ $R$ connected ∨ $R$ equivalence ⇔ $R$ universal.

***Corollaries***

1. Any constraint set containing any following pair of constraints is incoherent:
   (i) $R$ reflexive ∧ $R$ irreflexive.
   (ii) $R$ symmetric ∧ $R$ asymmetric.
2. (i) Any constraint set containing ($R$ asymmetric ∨ $R$ intransitive ∨ $R$ inEuclidean ∨ $R$ acyclic) ∧ $R$ reflexive is incoherent.
   (ii) Any constraint set containing $R$ acyclic ∧ $R$ symmetric is incoherent.
   (iii) Any constraint set containing ($R$ asymmetric ∨ $R$ intransitive ∨ $R$ acyclic) ∧ $R$ Euclidean is incoherent.
   (iv) Any constraint set containing $R$ transitive ∧ $R$ inEuclidean is incoherent.
   (v) Any constraint set containing $R$ connected ∧ ($R$ intransitive ∨ $R$ inEuclidean) is incoherent.
   (vi) Any constraint set containing ($R$ irreflexive ∨ $R$ asymmetric ∨ $R$ intransitive ∨ $R$ inEuclidean ∨ $R$ acyclic) ∧ $R$ equivalence is incoherent.
   (vii) Any constraint set containing ($R$ asymmetric ∨ $R$ intransitive ∨ $R$ inEuclidean ∨ $R$ acyclic) ∧ $R$ irreflexive is not minimal, as irreflexivity is redundant (i.e., (asymmetry ∨ intransitivity ∨ inEuclideanity ∨ acyclicity) ⇒ irreflexivity).
3. (i) Any constraint set containing ($R$ irreflexive ∨ $R$ asymmetric ∨ $R$ inEuclidean) ∧ $R$ acyclic is not minimal, as irreflexivity, asymmetry, and inEuclideanity are redundant (i.e., acyclicity ⇒ (irreflexivity ∧ asymmetry ∧ inEuclideanity)).
   (ii) Any constraint set containing $R$ symmetric ∧ $R$ Euclidean (as symmetry is redundant) or $R$ transitive ∧ $R$ Euclidean (as transitivity is redundant), or $R$ symmetric ∧ $R$ transitive ∧ $R$ Euclidean (as Euclideanity is redundant), or $R$ equivalence ∧ $R$ Euclidean ∧ $R$ reflexive or $R$ equivalence ∧ $R$ reflexive ∧ $R$ symmetric ∧ $R$ transitive (as either equivalence or reflexivity and Euclideanity, or reflexivity, symmetry, and transitivity, respectively, are redundant) is not minimal (i.e., Euclideanity ⇔ symmetry ∧ transitivity, Euclideanity ∧ reflexivity ⇔ equivalence, and reflexivity ∧ symmetry ∧ transitivity ⇔ equivalence).
4. (i) Any constraint set containing ($R$ reflexive ∨ $R$ symmetric ∨ $R$ Euclidean ∨ $R$ equivalence ∨ $R$ connected) ∧ $R$ transitive ∧ $R$ intransitive is incoherent.
   (ii) Any constraint set containing ($R$ irreflexive ∨ $R$ asymmetric ∨ $R$ inEuclidean) ∧ $R$ transitive ∧ $R$ intransitive is not minimal, as irreflexivity, asymmetry, and inEuclideanity are redundant (i.e., transitivity ∧ intransitivity ⇒ irreflexivity ∧ asymmetry ∧ inEuclideanity).
5. (i) Any constraint set containing $R$ Euclidean ∧ $R$ inEuclidean ∧ $R$ equivalence is incoherent.
   (ii) Any constraint set containing $R$ intransitive ∧ $R$ Euclidean ∧ $R$ inEuclidean is not minimal, as intransitivity is redundant (i.e., Euclideanity ∧ inEuclideanity ⇒ intransitivity).
6. (i) Any constraint set containing $R$ irreflexive ∧ $R$ symmetric ∧ $R$ transitive is incoherent.
   (ii) Any constraint set containing $R$ irreflexive ∧ $R$ asymmetric ∧ $R$ transitive is not minimal, as asymmetry is redundant (i.e., irreflexivity ∧ transitivity ⇒ asymmetry).
7. (i) Any constraint set containing $R$ symmetric ∧ $R$ intransitive ∧ $R$ inEuclidean is not minimal, as inEuclideanity is redundant (i.e., symmetry ∧ intransitivity ⇒ inEuclideanity).
   (ii) Any constraint set containing $R$ symmetric ∧ $R$ transitive ∧ $R$ intransitive ∧ $R$ Euclidean ∧ $R$ inEuclidean is not minimal, as Euclideanity and inEuclideanity are redundant (i.e., symmetry ∧ transitivity ∧ intransitivity ⇒ Euclideanity ∧ inEuclideanity).
8. (i) Any constraint set containing $R$ symmetric ∧ $R$ inEuclidean ∧ $R$ connected is incoherent.
   (ii) Any constraint set containing $R$ symmetric ∧ $R$ intransitive ∧ $R$ inEuclidean is not minimal, as intransitivity is redundant (i.e., symmetry ∧ inEuclideanity ⇒ intransitivity).

(iii) Any constraint set containing *R* symmetric ∧ *R* transitive ∧ *R* intransitive ∧ *R* Euclidean ∧ *R* inEuclidean is not minimal, as transitivity and intransitivity are redundant (i.e., symmetry ∧ Euclideanity ∧ inEuclideanity ⇒ transitivity ∧ intransitivity).

9. Any constraint set containing *R* asymmetric ∧ *R* transitive ∧ *R* acyclic is not minimal, as acyclicity is redundant (i.e., asymmetry ∧ transitivity ⇒ acyclicity).
10. (i) Any constraint set containing *R* symmetric ∧ *R* intransitive ∧ *R* connected is incoherent.
(ii) Any constraint set containing *R* symmetric ∧ *R* Euclidean ∧ *R* connected is not minimal, as Euclideanity is redundant (i.e., symmetry ∧ connectivity ⇒ Euclideanity).
(iii) Any constraint set containing *R* symmetric ∧ *R* transitive ∧ *R* connected is not minimal, as transitivity is redundant (i.e., symmetry ∧ connectivity ⇒ transitivity).
(iv) Any constraint set containing *R* reflexive ∧ *R* symmetric ∧ *R* connected ∧ *R* equivalence is not minimal, as equivalence is redundant (i.e., reflexivity ∧ symmetry ∧ connectivity ⇒ equivalence).
11. (i) Any constraint set containing *R* intransitive ∧ *R* Euclidean and not containing *R* inEuclidean is incoherent.
(ii) Any constraint set containing *R* acyclic ∧ *R* Euclidean and not containing *R* inEuclidean is incoherent.
(iii) Any constraint set containing *R* connected ∧ *R* inEuclidean and not containing *R* Euclidean is incoherent.
12. Any constraint set containing *R* transitive ∧ *R* inEuclidean ∧ *R* connected is incoherent.
13. Any constraint set containing *R* intransitive ∧ *R* Euclidean ∧ *R* inEuclidean is not minimal, as inEuclideanity is redundant (i.e., intransitivity ∧ Euclideanity ⇒ inEuclideanity).
14. Any constraint set containing *R* transitive ∧ *R* inEuclidean ∧ *R* connected is incoherent.
15. Any constraint set containing *R* transitive ∧ *R* acyclic ∧ *R* connected is not minimal, as transitivity is redundant (i.e., acyclicity ∧ connectivity ⇒ transitivity).
16. Any constraint set containing *R* intransitive ∧ *R* acyclic ∧ *R* connected is incoherent.
17. Any constraint set containing *R* inEuclidean ∧ *R* connected ∧ (*R* asymmetric ∨ *R* acyclic) is not minimal, as asymmetry and acyclicity are redundant (i.e., inEuclideanity ∧ connectivity ⇒ asymmetry ∧ acyclicity).
18. Any constraint set containing *R* connected ∧ *R* inEuclidean ∧ *R* acyclic is not minimal, as either inEuclideanity or acyclicity are redundant (i.e., inEuclideanity ∧ connectivity ⇒ acyclicity and acyclicity ∧ connectivity ⇒ inEuclideanity).
19. Any constraint set containing *R* connected ∧ *R* reflexive ∧ (*R* symmetric ∨ *R* Euclidean) ∨ *R* connected ∧ *R* equivalence is not fundamental, as it is universal (i.e., reflexivity ∧ connectivity ∧ (symmetry ∨ Euclideanity) ∨ connected ∧ equivalence ⇒ universality), and should instead be stored as a query/view computing $S^2$.

*MatBase* stores in its metacatalog these 19 above corollaries in four tables presented in the following subsections.

***Table COROLARIES***

Table *COROLARIES* (see Figure 2) stores data about the corollaries on the coherence and minimality of constraint sets (a surrogate primary autogenerated key x, corollaries' types, names, bodies, book volume, subsection, and page number in which they appear in [8], etc.). *COROLARIES* also stores data for all other 65 (E)MDM constraint types [1], not only for the 11 dyadic relation ones. Data from this table (which was manually entered) is used for providing users with context-sensitive questions, warnings, and error messages.

***Table DRCCoherencies***

Table *DRCCoherencies* (see Figure 3) stores data about the coherency of the non-trivial dyadic relationship combinations (out of the $2^{11}$ - 1 = 2,047 possible ones). Abbreviations of the 12 columns of *DRCCoherencies* after the primary key *x* have the following meanings: *Ch* = Coherent?, *C* = Connected?, *A* = Acyclic?, *Q* = eQuivalence?, *IE* = InEuclidean?, *E* = Euclidean?, *IT* = InTransitive? , *T* = Transitive?, *AS* = Asymmetric?, *S* = Symmetric? *IR* = Irreflexive?, *R* = Reflexive?.

The unique combination numbers x are computed as the decimal equivalents of the corresponding binary ones, just like for all other tables storing constraint type combinations (where *C* is multiplied by $2^{10}$ = 1024, *A* by $2^9$ = 512, …, and *R* by $2^0$ = 1, i.e., x = [R]+2*[IR]+4*[S]+8*[AS]+16*[T]+32*[IT]+64*[E]+128*[IE]+256*[Q]+512*[A]+1024*[C]).

For example, combinations {ASymmetric} and {ASymmetric, inEuclidean} have 8 and 136, respectively, as values for $x$ (ASymmetric being multiplied by $2^3$ and inEuclidean by $2^7$) and are coherent, while the one for $x = 41$, i.e., {InTransitive, ASymmetric, Reflexive} is incoherent (as, according to Corollary 2($i$), any intransitive or/and asymmetric dyadic relation cannot be reflexive as well).

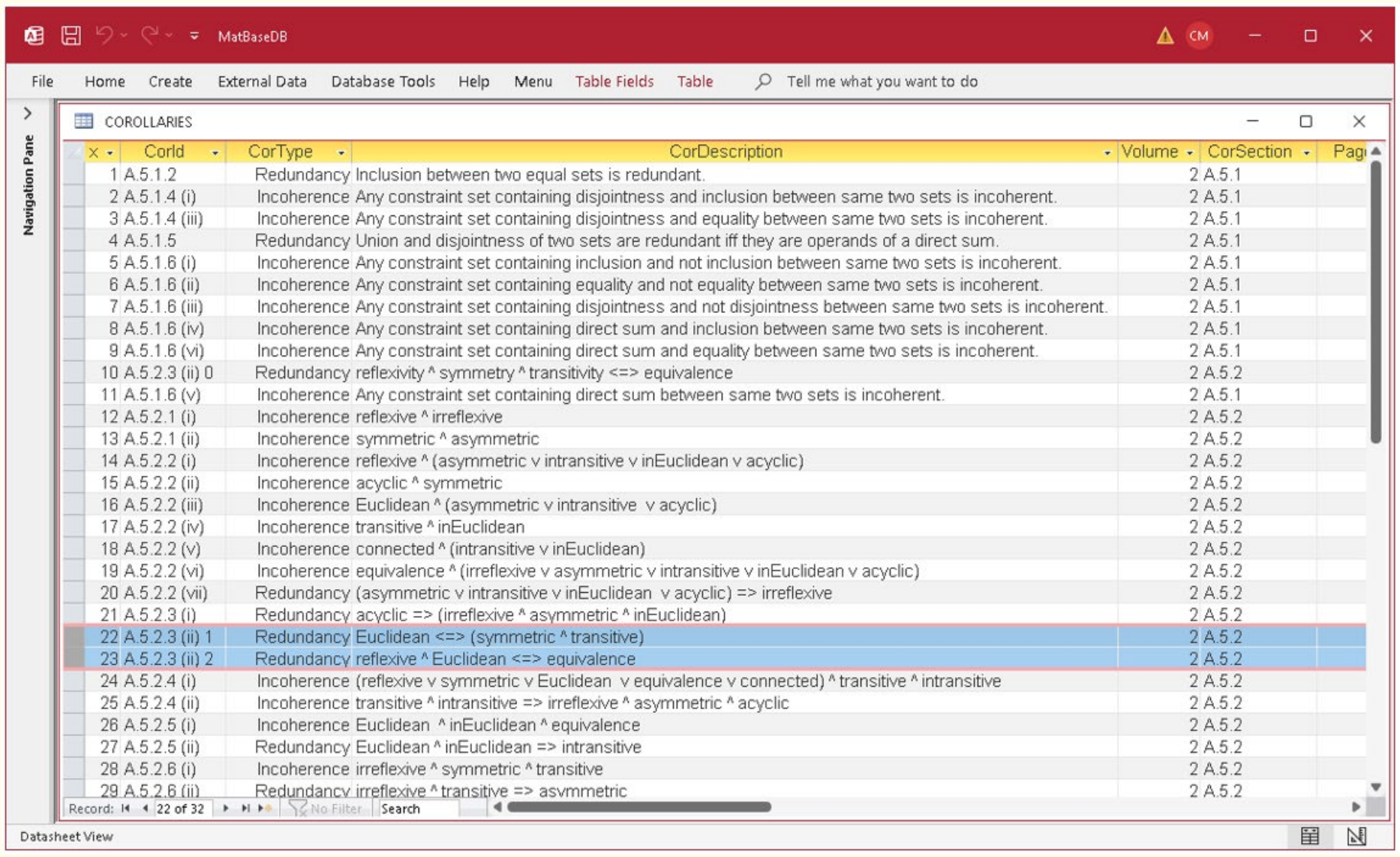

***Figure 2:*** MS Access *MatBase COROLLARIES* table for storing corollaries on the coherence and minimality of constraint sets.

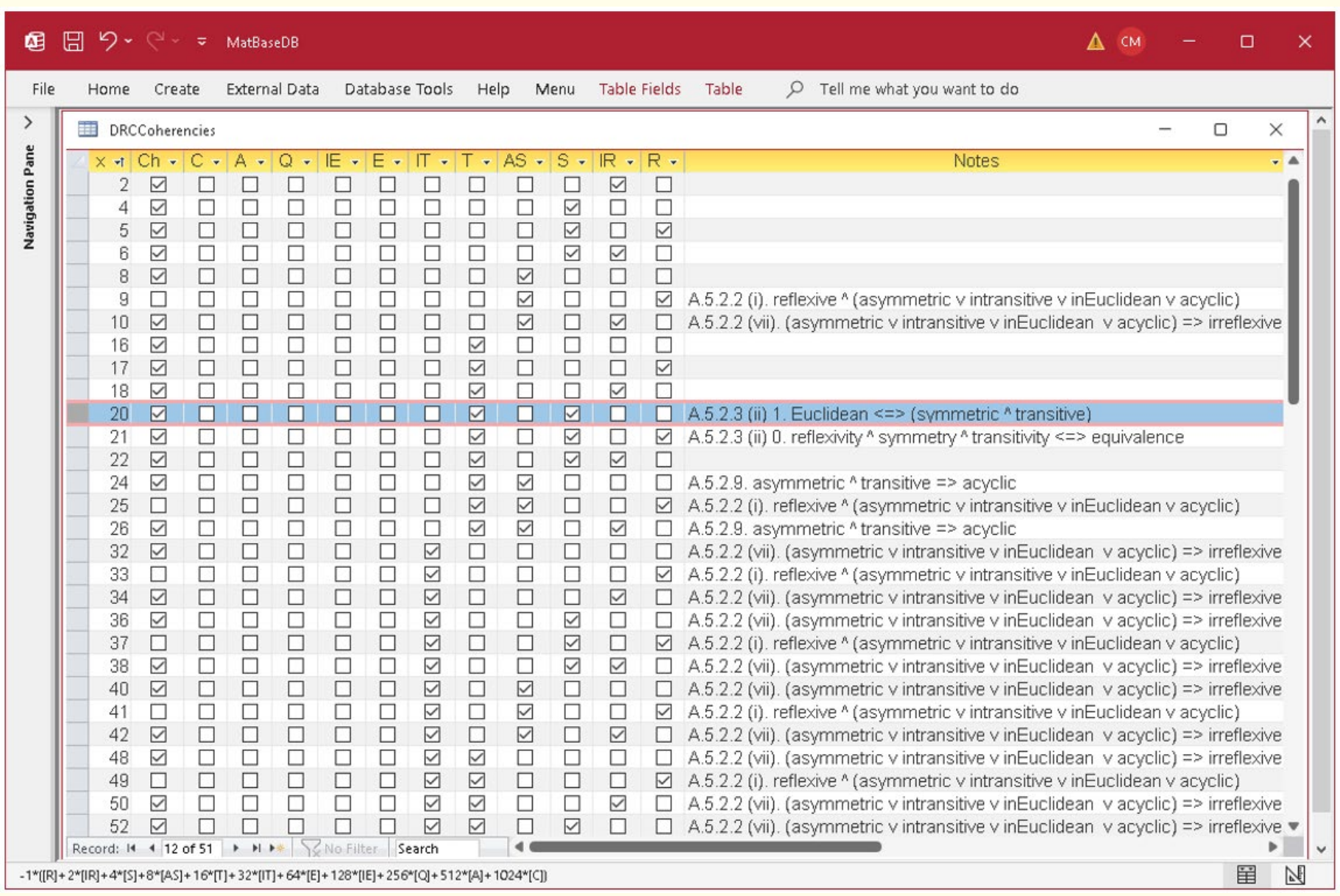

***Figure 3:*** MS Access *MatBase DRCCoherencies* table for storing non-trivial combinations of dyadic relation constraint types.

Obviously, *Notes* is a foreign key referencing the primary key *x* of table *COROLLARIES*, from which its combo-box displays the corresponding values from the *CorId* and *CorDescription* columns for incoherent and not minimal combinations. The corresponding combo-box row source SQL statement is the following:

SELECT x, CorId & ". " & CorDescription AS [CorollaryID, Body] FROM COROLLARIES

WHERE CorSection="A.5.2" ORDER BY CorId & ". " & CorDescription;

*DRCCoherencies* instance was automatically generated using SQL insert and update queries as follows: a query first inserted all non-trivial possible combinations (the trivial ones, i.e., those from Corollary 1, are not stored); then, queries were run for each of the other 18 incoherence results, marking corresponding combinations as incoherent. For example, the query corresponding to Corollary 2(*i*) is the following one:

UPDATE [DRCCoherencies] SET [Ch] = False, [Notes] = 14 WHERE [R] AND ([AS] OR [IT] OR [IE] OR [A]);

Finally, queries were run for all redundancy corollaries to update *notes* for the coherent but not minimal constraint set ones. For example, the query corresponding to Corollary 2(*vii*) is the following:

UPDATE [DRCCoherencies] SET [Notes] = 20 WHERE [Ch] AND ([AS] OR [IT] OR [IE] OR [A]);

Generally, more than one redundancy corollary may apply to a constraint set. For example, the set {Reflexive, Symmetric, Transitive, Euclidean, Equivalence} has Equivalence redundant according to corollary A.5.2.3 (*ii*) 0 (see row 10 from Figure 2) but also to corollary A.5.2.3 (*ii*) 2 (see row 23 from Figure 2), as well as Euclidean, according to corollary A.5.2.3 (*ii*) 1 (see row 22 from Figure 2). Consequently, there is also a table *DRCAdditionalRedund* in the metacatalog of *MatBase* for storing the rest of redundancies for combinations having more than one; its structure is identical to the one of the table *DRCRedundancies* presented in the next subsection and its instance is also automatically populated with SQL INSERT statements. As this table is used only for automatically adding rows to the *DRCRedundancies* table and then for displaying accurate context-sensitive information and error messages, to keep things simple in this paper we are not providing more details on how it is used.

***Table DRCRedundancies***

Table *DRCRedundancies* (see Figure 4) stores data on the minimality of dyadic relation constraint sets. Column *DRCCombination* is a foreign key referencing the primary key *x* of table *DRCCoherencies*; column *Notes* is absolutely similar to the homonym one in table *DRCCoherencies*, except for the fact that it points to the subset of corollaries having type "Redundancy" (so in the corresponding combo-box row source SQL statement WHERE clause there is a CorType atom as well *and*-ed with the other ones); finally, the column *Redundancy* stores the redundant constraint types that make the corresponding constraint sets not minimal.

As an example, for a constraint set having *DRCCombination* = 65, which corresponds in table *DRCCoherencies* to the line having x = 65, which encodes a set of type {Euclidean, Reflexive} (as Euclideanity is multiplied by $2^6$ and reflexivity by $2^0$), there are three rows in table *DRCRedundancies* (see the selected rows from Figure 4) storing the fact that both symmetry and transitivity (according to the corollary A.5.2.3 (*ii*) 1, see row 22 from Figure 2), as well as equivalence (according to the corollary A.5.2.3 (*ii*) 2, see row 23 from Figure 2) are to be added as redundant ones to such constraint sets.

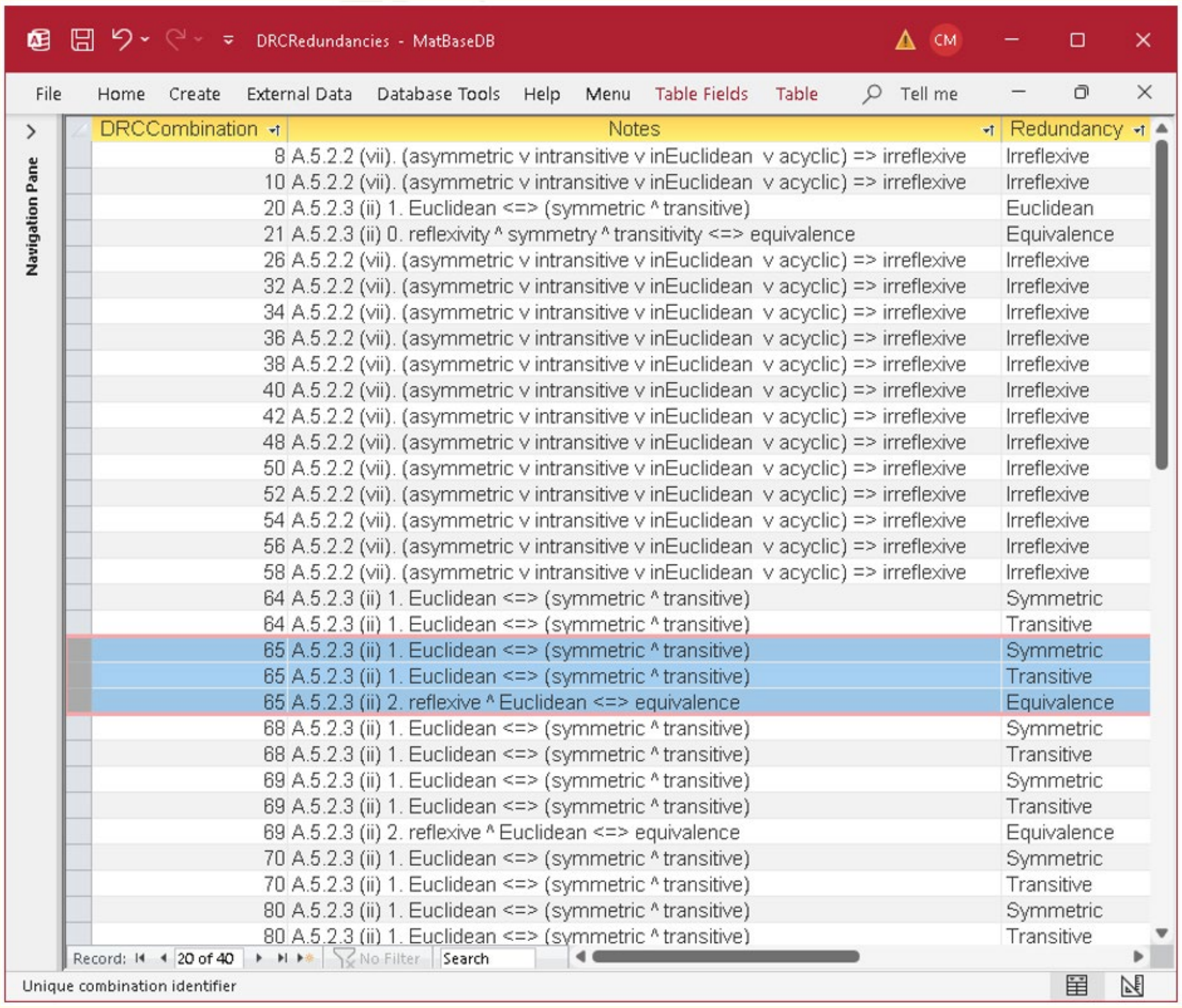

| DRCCombination | Notes | Redundancy |
|---|---|---|
| 8 | A.5.2.2 (vii). (asymmetric v intransitive v inEuclidean v acyclic) => irreflexive | Irreflexive |
| 10 | A.5.2.2 (vii). (asymmetric v intransitive v inEuclidean v acyclic) => irreflexive | Irreflexive |
| 20 | A.5.2.3 (ii) 1. Euclidean <=> (symmetric ^ transitive) | Euclidean |
| 21 | A.5.2.3 (ii) 0. reflexivity ^ symmetry ^ transitivity <=> equivalence | Equivalence |
| 26 | A.5.2.2 (vii). (asymmetric v intransitive v inEuclidean v acyclic) => irreflexive | Irreflexive |
| 32 | A.5.2.2 (vii). (asymmetric v intransitive v inEuclidean v acyclic) => irreflexive | Irreflexive |
| 34 | A.5.2.2 (vii). (asymmetric v intransitive v inEuclidean v acyclic) => irreflexive | Irreflexive |
| 36 | A.5.2.2 (vii). (asymmetric v intransitive v inEuclidean v acyclic) => irreflexive | Irreflexive |
| 38 | A.5.2.2 (vii). (asymmetric v intransitive v inEuclidean v acyclic) => irreflexive | Irreflexive |
| 40 | A.5.2.2 (vii). (asymmetric v intransitive v inEuclidean v acyclic) => irreflexive | Irreflexive |
| 42 | A.5.2.2 (vii). (asymmetric v intransitive v inEuclidean v acyclic) => irreflexive | Irreflexive |
| 48 | A.5.2.2 (vii). (asymmetric v intransitive v inEuclidean v acyclic) => irreflexive | Irreflexive |
| 50 | A.5.2.2 (vii). (asymmetric v intransitive v inEuclidean v acyclic) => irreflexive | Irreflexive |
| 52 | A.5.2.2 (vii). (asymmetric v intransitive v inEuclidean v acyclic) => irreflexive | Irreflexive |
| 54 | A.5.2.2 (vii). (asymmetric v intransitive v inEuclidean v acyclic) => irreflexive | Irreflexive |
| 56 | A.5.2.2 (vii). (asymmetric v intransitive v inEuclidean v acyclic) => irreflexive | Irreflexive |
| 58 | A.5.2.2 (vii). (asymmetric v intransitive v inEuclidean v acyclic) => irreflexive | Irreflexive |
| 64 | A.5.2.3 (ii) 1. Euclidean <=> (symmetric ^ transitive) | Symmetric |
| 64 | A.5.2.3 (ii) 1. Euclidean <=> (symmetric ^ transitive) | Transitive |
| 65 | A.5.2.3 (ii) 1. Euclidean <=> (symmetric ^ transitive) | Symmetric |
| 65 | A.5.2.3 (ii) 1. Euclidean <=> (symmetric ^ transitive) | Transitive |
| 65 | A.5.2.3 (ii) 2. reflexive ^ Euclidean <=> equivalence | Equivalence |
| 68 | A.5.2.3 (ii) 1. Euclidean <=> (symmetric ^ transitive) | Symmetric |
| 68 | A.5.2.3 (ii) 1. Euclidean <=> (symmetric ^ transitive) | Transitive |
| 69 | A.5.2.3 (ii) 1. Euclidean <=> (symmetric ^ transitive) | Symmetric |
| 69 | A.5.2.3 (ii) 1. Euclidean <=> (symmetric ^ transitive) | Transitive |
| 69 | A.5.2.3 (ii) 2. reflexive ^ Euclidean <=> equivalence | Equivalence |
| 70 | A.5.2.3 (ii) 1. Euclidean <=> (symmetric ^ transitive) | Symmetric |
| 70 | A.5.2.3 (ii) 1. Euclidean <=> (symmetric ^ transitive) | Transitive |
| 80 | A.5.2.3 (ii) 1. Euclidean <=> (symmetric ^ transitive) | Symmetric |
| 80 | A.5.2.3 (ii) 1. Euclidean <=> (symmetric ^ transitive) | Transitive |

***Figure 4:*** MS Access *MatBase DRCRedundancies* table for storing redundant constraint types of not minimal dyadic relation constraint sets.

The instance of *DRCRedundancies* was also automatically generated by running SQL queries for each redundancy corollary. For example, for corollary A.5.2.3 (*ii*) 1 (see row 22 from Figure 2 and, e.g., row 20 from Figure 3), the following two SQL statements were run for inserting the first two selected lines from Figure 4:

INSERT INTO DRCRedundancies (DRCCombination, [Notes], Redundancy)

SELECT x, [Notes], "S" FROM DRCCoherencies WHERE [Ch] AND [E];

INSERT INTO DRCRedundancies (DRCCombination, [Notes], Redundancy)

SELECT x, [Notes], "T" FROM DRCCoherencies WHERE [Ch] AND [E];

Please note that these SQL statements need to be recursively run up until no new redundancy is added to DRCRedundancies. For example, let us assume that a SQL statement called $Q_1$ is run first for adding acyclicity to all constraint sets containing asymmetry and transitivity; obviously, no row would be inserted for the set {Irreflexive, Transitive}; then, assume that a SQL statement called $Q_2$ is run afterwards for adding asymmetry as redundant to all constraint sets containing irreflexivity and transitivity; now, the set {Irreflexive, Transitive} becomes {Irreflexive, Transitive, Asymmetric}; obviously, $Q_1$ must be run one more time for adding acyclicity as well to this set.

***MatBase Algorithm DRCSCMEA***

When a user tries to add a new constraint c to a dyadic relation *R* (over a set *S* and having an associated constraint set *C*) by clicking the corresponding checkbox shown in Figure 1, *MatBase* is first computing the *x* value of this new constraint set and is looking for it in table *DRCCoherencies*. If it doesn't find it, which only occurs when this set also contains the negation of c (e.g., *c* = Reflexivity and *C* contains Irreflexivity), then it unchecks the corresponding box and displays the appropriate error message. If it finds it but corresponding to an incoherent combination, it rejects it as well, similarly. If the corresponding combination is coherent but has in table *DRCRedundancies* a Universal redundancy (i.e., the new constraint set corresponds to a universal relation), then it informs the user about it and asks him/her whether replacing *R* with a view computing $S^2$ is desired; if the answer is Yes, then it proceeds accordingly, otherwise it automatically unchecks the corresponding box. Finally, in all other cases it checks whether the current *R*'s data instance satisfies *c* and if this is not the case it rejects *c* as well, similarly to the above cases.

Whenever *c* is accepted both syntactically (i.e., from the coherence point of view) and semantically (i.e., from the data satisfiability one), *MatBase* adds to *R*'s form programming class (automatically generated immediately after table *R* has been added to the current db) calls to the corresponding *c* enforcement methods (which are publicly stored in its *Constraint* library) [12]. Moreover, if formerly not redundant constraints have become redundant, *MatBase* deletes from the *R*'s form programming class the code calling the corresponding public enforcement methods. Finally, it also automatically checks all newly redundant constraints, according to *DRCRedundancies* data for the newly *x* value from *DRCCoherencies* (e.g., if *C* = {Transitive}, *c* = Asymmetric, then *C*' = {Transitive, Asymmetric, Acyclic, Irreflexive}, with Acyclic and Irreflexive being both redundant).

When a user tries to remove a constraint *c* by unchecking its corresponding box shown in Figure 1, *MatBase* first computes the *x* value for the initial associated constraint set *C* and looks for *c* in *DRCRedundancies* table for *x*; if it finds it, then rejects the deletion attempt (as redundant constraints may not be deleted); otherwise, it removes from the *R*'s form programming class the calls to the constraint enforcement methods corresponding to *c*, then computes the corresponding new x value for *C*' and, finally, unchecks all formerly redundant constraints that are not implied anymore (e.g., if *C* = {Transitive, Asymmetric, Acyclic, Irreflexive} and *c* = Transitive is deleted from it, then Acyclic is also deleted and *C*' = {Asymmetric, Irreflexive}; note that irreflexivity is not deleted as well, because it remains implied by asymmetry).

Figure 5 presents the corresponding pseudocode algorithm used by *MatBase* to enforce dyadic relation constraints, while guaranteeing the satisfiability, coherency, and minimality of such constraint sets.

## Results and Discussion

***Proposition 21.***

Algorithm *DRCSCMEA* from Figure 5 has the following properties:

(i) its complexity is a constant (i.e., $O(k)$).

(ii) it guarantees the satisfiability, coherence, and minimality of dyadic relation constraint sets.

(iii) it is solid, complete, and optimal.

```
ALGORITHM DRCSCMEA (Dyadic Relation Constraint Sets Satisfiability, Coherence, and Minimality Enforcement)
Input: the current (E)MDM scheme constraint set C for a dyadic relationship R over a set S and a constraint type c ∈
{reflexive, irreflexive, symmetric, asymmetric, transitive, intransitive, Euclidean, inEuclidean, equivalence, acyclic,
connected} to be added to or removed from C.

Output: the updated coherent and minimal (E)MDM scheme C' for R.

Strategy:

C' = C;
if c is to be removed from C then compute the x value for C;
  if c's type is found redundant in a DRCRedundancy row having DRCCombination = x then
    display c & " cannot be removed as it is implied by other constraints, according to " & Notes(x);
  else C' = C' – { c };
     remove from the R's class the calls to the constraint enforcement methods corresponding to c;
     compute the corresponding new x value for C';
     remove from C' all formerly redundant constraints that are not implied anymore;
  end if;
else        // c is to be added to C
  compute the corresponding x value for C' ∪ c;
  if ¬Ch(x) then display c & " cannot be added, as, according to " & Notes(x) & ", the constraint set of " & R &
                          " would become incoherent!";
  elseif x is missing from DRCCoherencies then display c & " cannot be added, as the constraint set of " & R &
                                                             "implies ¬" & c & ", so it would become incoherent!";
  elseif DRCRedundancy has a row having DRCCombination = x and Redundancy = "Universal" then
    display "Adding " & c & " to the current constraint set of " & R & " would make it universal, according to " & Notes(x)
    display "Are you sure you want MatBase to drop " & R & " and replace it with a view having same name and " &
           "computing " & S & " × " & S & "?";
    if the answer is Yes then drop R; add to the db schema a view R computing S × S;  end if;
  else
    if c is not satisfied by the current db instance then display c & " cannot be added to the constraint set of " & R &
                                                                   ", as its current instance does not satisfy it!"
    else C' = C' ∪ { c };
       add to the R's class the calls to the corresponding constraint type enforcement methods;
       mark as true all redundant constraints that are implied by C'according to DRCRedundancies;
       remove from the R's class enforcement code for any constraint that just became redundant;
    end if;
  end if;
end if;
End ALGORITHM DRCSCMEA;
```

***Figure 5:*** *MatBase* pseudocode Algorithm *DRCSCMEA*.

*Proof*

(i) Trivially, it does not contain any loop, so it always ends in finite time after a (small) number of finite steps.

(ii) (*satisfiability*) Trivially, any void constraint set is satisfied by any data instance of any dyadic relation and any non-void constraint set that is satisfied by a data instance remains satisfied after removing one of its constraints; as *DRCSCMEA* does not accept adding a new constraint to the constraint set of a dyadic relation if its instance does not satisfy it as well, it follows, obviously, that *DRCSCMEA* guarantees the satisfiability of such constraint sets.

(*coherence*) Trivially, any void constraint set is coherent and any non-void coherent constraint set remains coherent after removing one of its constraints; as *DRCSCMEA* does not accept adding a new constraint to the constraint set of a dyadic relation if this would result in an incoherent set, it follows, obviously, that *DRCSCMEA* guarantees the coherence of such constraint sets as well.

(*minimality*) Trivially, any void constraint set is minimal; as *DRCSCMEA* is never enforcing redundant constraints but only signals them to the users for their info and is recomputing the subset of redundant constraints after accepting both adding and deleting a constraint, it follows, obviously, that *DRCSCMEA* guarantees the minimality of such constraint sets as well.

(iii) (*solidity*) Trivially, *DRCSCMEA* accepts to add to or delete from dyadic relation constraint sets only dyadic relation constraint types.

(*completeness*) Trivially, *DRCSCMEA* accepts to add to or delete from dyadic relation constraint sets all types of dyadic relation constraints.

(*optimality*) Trivially, *DRCSCMEA* manages satisfiable, coherent, and minimal dyadic relation constraint sets in the minimum possible number of steps, with the minimum possible accesses to the 3 tables presented in the previous section (and which are stored on external disks). *Q.E.D.*

The actual corresponding algorithms (written both in MS VBA and .NET C# with embedded SQL, respectively) are a little bit more complex, both to gain execution speed (by avoiding unnecessary disk reads), to prevent users from making unwanted mistakes, and to provide maximum possible accuracy for the context-sensitive messages it displays. For example, whenever the current dyadic relation *R* has no constraints and the user adds one, it accepts it immediately if the current *R*'s instance satisfies it, as there may not be any corresponding either incoherency or redundancy. For example, whenever the user unchecks a constraint box, even if the corresponding deletion is possible *MatBase* displays a deletion confirmation message, does not proceed with the deletion if the request is not confirmed, and automatically undoes unchecking of the corresponding box. Moreover, if the request is confirmed and *c* is the only constraint of *C*, *MatBase* does not search for newly redundant constraints, as none may exist.

Obviously, the ultimate goal of the design and development of dbs and db software applications is to provide customers, first of all, with the tools that are not only user-friendly, but, above all, guaranteeing the highest possible data quality for their dbs and information extracted from them. If these tools do not guarantee the satisfiability and coherence of the associated constraint sets (be them enforced at the db or/and at the db software application levels), then junk data might (accidentally or purposely, it does not matter) be stored in their dbs, which leads to junk information extracted from them. Moreover, if these constraint sets are not minimal (which, yes, does not impact data quality), then the corresponding db software applications run unnecessarily slower, to the dissatisfaction of their customers.

## Conclusion

We provided concise but accurate mathematical definitions for dyadic relations, their properties viewed as constraint types from the db perspective, as well as for the satisfiability, coherence, and minimality of such constraint sets.

We presented and discussed the pseudocode algorithm used by *MatBase*, our intelligent DBMS prototype based on both the relational, entity-relationship data models, as well as on our (E)MDM (which incorporates the dyadic relation constraint types), for enforcing dyadic relation constraints, by guaranteeing the satisfiability, coherence, and minimality of such constraint sets, also including description of the tables from its metacatalog needed for managing the corresponding metadata.

We proved that this algorithm actually guarantees both satisfiability, coherence, and minimality, while being fast, solid, complete, and optimal.

Moreover, this paper also proves the formidable power of using mathematics (in particular, the naïve theory of sets, relations and functions coupled with the first-order predicate calculus with equality) in dbs and db software applications design and development.

## Conflict of interest

The author declare that the research was conducted in the absence of any commercial or financial relationships that could be construed as a potential conflict of interest.

## Acknowledgements

This research was not sponsored by anybody and nobody other than its author contributed to it.